\documentclass[a4paper,12pt]{article}

\usepackage{chemformula}
\usepackage{amssymb}
\usepackage{amsmath}
\usepackage{amsfonts}
\usepackage{amsthm}

\newtheorem{theorem}{Theorem}

\DeclareMathOperator{\End}{End}
\DeclareMathOperator{\grad}{grad}
\DeclareMathOperator{\diver}{div}

\DeclareMathOperator{\Tr}{Tr}

\newcommand{\reals}{\mathbb{R}}

\newcommand{\realproj}[1]{\reals\mathbb{P}^{#1}}
\newcommand{\sphere}[1]{\mathbb{S}^{#1}}

\newcommand{\LieDerivative}[2]{\mathcal{L}_{#1}{#2}}
\newcommand{\matD}[1]{\frac{{d}}{{d}t}{#1}}
\newcommand{\mnfld}{M}
\newcommand{\mnfldbase}{B}
\newcommand{\metric}{g}

\newcommand{\metricM}{g_{\mnfld}}
\newcommand{\metricB}{g_{\mnfldbase}}
\newcommand{\metricF}{g_F}
\newcommand{\conn}[1]{\nabla_{#1}}
\newcommand{\LCconn}{\nabla}
\newcommand{\covdif}{d_{\LCconn}}
\newcommand{\tanspace}{T}
\newcommand{\cotanspace}{T^{*}}
\newcommand{\cnj}[1]{{#1}^{\prime}}
\newcommand{\idop}{\mathbf{1}}

\newcommand{\parder}[2]{\frac{\partial #1}{\partial #2}}
\newcommand{\inner}[2]{{\left(#1,#2\right)}}

\newcommand{\contactForm}{\alpha}

\newcommand{\contactSpace}{\Phi}
\newcommand{\symplecticSpace}{\widetilde\contactSpace}
\newcommand{\flowvel}{{X}}
\newcommand{\stress}{\sigma}

\newcommand{\dfrm}{\Delta}

\newcommand{\entr}{s}

\newcommand{\temp}{\theta}
\newcommand{\dens}{\rho}
\newcommand{\intenergy}{\varepsilon}
\newcommand{\totenergy}{e}
\newcommand{\helm}{h}
\newcommand{\energyFlux}{\mathcal{J}_e}
\newcommand{\heatFlow}{\mathcal{J}_q}

\newcommand{\thermcond}{\chi}

\title{Continuum mechanics of media with inner structures}

\author{Anna Duyunova\\
V. A. Trapeznikov Institute\\
of
Control Sciences\\
of Russian Academy
of Sciences,\\
Moscow, Russia\\
\texttt{anna.duyunova@yahoo.com}
\and
Valentin Lychagin\\
V. A. Trapeznikov Institute\\
of
Control Sciences\\
of Russian Academy
of Sciences,\\
Moscow, Russia\\
University of Troms\o, Norway\\
\texttt{valentin.lychagin@uit.no}
\and
Sergey Tychkov\\
V. A. Trapeznikov Institute\\
of
Control Sciences\\
of Russian Academy
of Sciences,\\
Moscow, Russia\\
\texttt{sergey.lab06@ya.ru}
}

\begin{document}
\maketitle
\begin{abstract}
	We propose a geometrical approach to the mechanics
	of continuous media equipped
	with inner structures and give the basic
	(mass conservation, Navier-Stokes
	and energy conservation) equations of their motion.
	\end{abstract}
\section{Introduction}

The classical continuum mechanics deals with media in
three-dimen\-si\-onal spaces
and describes them by a system of partial differential equations on
a 3D manifold $\mnfld$. Usually, $\mnfld$ is
a domain in $\reals^3$ or $\reals^{+}\times\sphere{2}$. In
such cases, elements of the medium are points. The idea that the medium
has inner structure leads to a more interesting picture.

Let us consider, for example, \textit{a diatomic gas}. A molecule of such
gas is composed of two atoms. We have \textit{a homonuclear gas} if its
molecule is composed of two atoms of the same element, e.g. oxygen \ch{O2},
or \textit{a heteronuclear gas} if it is composed of two atoms of different
elements, like carbon monoxide \ch{CO}. Thus, for description of such gases, we
use $\mnfld=\reals^{3}\times\realproj{2}$ for the homonuclear gases and
$\mnfld=\reals^{3}\times\sphere{2}$ for heteronuclear ones. In the case
of atmospheric gases, the configuration manifold $\mnfld$ is the total space of the
bundle $\pi :M\rightarrow \reals^{+}\times\sphere{2}$, with fibers
diffeomorphic to $\realproj{2}$ or $\sphere{2}$, depending on the
type of the gas. Here the projection $\pi$ indicates the mass center of
molecule. The case of water, \ch{H2O}, is more interesting, here the
configuration space of inner states is the circle bundle of the tangent
space to a two-dimensional sphere, and this space is
diffeomorphic to $\realproj{3}$ (see, for example, \cite{Kr}).
An example of a medium composed with solids,
i.~e. the bundle
$\pi : \reals^3\times SO(3)\rightarrow\reals^3$ is trivial,
were given by Cosserat brothers \cite{Cos},
for a modern discussion see \cite{Vard}.

These examples give motivation for the following definition of \textit{a medium
with inner structure} used in this paper.

Namely, by a configuration space of such medium we mean:
\begin{enumerate}
\item A smooth bundle $\pi : \mnfld\rightarrow \mnfldbase$,
where $\mnfld$ and $\mnfldbase$ are Riemannian manifolds
equipped with metrics $\metricM$ and $\metricB$ respectively.
\item In order to compare inner structures (fibers of $\pi$)
at different points of $\mnfldbase$, we assume that the bundle $\pi$ is equipped with a connection $\conn{\pi}$.
This connection splits tangent spaces $T_{m}M$ into the vertical 
$T_{m}^{v}M$ and horizontal
$H_{m}\overset{\pi_{\ast}}{\simeq}T_{\pi(m)}B$ parts, i.~e.
$T_{m}M=T_{m}^{v}M\oplus H_{m}$,
where $T_{m}^{v}M$ and $H_{m}$ are $g_{M}$-orthogonal,
and the restriction of the metric $g_{M}$ on $H_{m}$
coincides with $g_{B}$. Moreover, we require that
the parallel transport of fibers along the
connection $\conn{\pi}$ be an isometry of fibers
with respect to the metric induced by $\metricM$.
\item Flow in the medium is given by a
$\pi$-projectable vector field
$\flowvel$ on $\mnfld$,
i.~e., $\flowvel$ preserves the bundle $\pi$.
This field can be split due
to the connection $\conn{\pi}$ into
the sum $\flowvel=\flowvel_H+\flowvel_V$,
where $\flowvel_V$ is a $\pi$-vertical field
and $\flowvel_H$ is a horizontal lift of the vector field
$\flowvel^{\prime}=\pi_{\ast}(\flowvel)$
on the base manifold $\mnfldbase$.
\end{enumerate}

The paper is organized as follows. First of all, we discuss the
thermodynamics of media based on the measurement \cite{LM}
of internal energy and
deformation tensor. Especially, we discuss Newtonian media
that have $O(\metricM)$ in a general case and
$O(\metricB)\times O(\metricF)$ symmetries,
when media have an inner structure.
Using the Rosenlicht and Procesi theorems (see \cite{Ros},
\cite{KrL},\cite{Proc}),
we find the fields of rational invariants,
and describe state
equations for internal energy and stress tensor in their terms.
We formulate the
thermodynamic state equations in terms of Lagrangian manifolds,
additionally equipped with a Riemannian structure.
This approach introduces these
manifolds as an intrinsic part of the
continuum mechanics, as well as it allows
us to describe different critical phenomena
in a pure geometrical manner (see,
for example, \cite{DLT1}, \cite{DLT2}, \cite{GLTR},
\cite{LR}). The second part
of the paper contains basic equations for
media: the mass conservation,
Navier-Stokes and energy conservation \cite{B}.
We outline shortly the
coordinateless method of getting these equations, which makes them
more transparent (at least from our point of view).

\section{Thermodynamics of media with inner structure}

The thermodynamics is based on measurement \cite{LM}
of two \textit{extensive
quantities}: the internal energy density $\intenergy$
and \textit{the rate of deformation}
$\dfrm=\covdif\flowvel\in\End\tanspace$,
where $\LCconn$ is the Levi-Civita connection associated with the metric
$g_{M},$ and $\covdif$ is the covariant differential.
The corresponding dual, or 
\textit{intensive,} quantities are
the temperature $\temp$ and the stress
tensor $\stress\in\End\cotanspace$. We will use
the duality of $\End\tanspace$ and $\End\cotanspace$
given by the pairing
$\left\langle\stress,\dfrm\right\rangle
=\Tr\,\stress\dfrm^*$.

Thus, the thermodynamic phase space of the medium is 
\[
\contactSpace=\reals^{3}\times
\End\cotanspace\times
\End\tanspace,
\]
with elements 
\[
(\entr,\temp,\intenergy,\stress,\dfrm),
\]
where $\entr$ is the entropy density.

In order to write down the first law of thermodynamics,
we adopt the following construction.

Let $(\mathcal{A},\star)$ be an $\reals$-algebra.
Consider the algebra $\mathcal{A}\otimes\End\tanspace$
(in this section $\otimes_{\reals}=\otimes$)
with the multiplication given by
$(a\otimes A)(b\otimes B)=
(a\star b)\otimes (AB)$. On this algebra, we define
a mapping called trace
$\Tr_{\mathcal{A}}: a\otimes A \mapsto a\Tr A$.
In the same way we define the algebra
$\mathcal{A}\otimes \End\cotanspace$. Then 
we define $\star$-pairing
$(\mathcal{A}\otimes \End\cotanspace)\times
(\mathcal{A}\otimes \End\tanspace) \rightarrow\mathcal{A}$ as follows 
\[
(a\otimes A)\star (b\otimes B)=
(a\star b)\Tr_{\mathcal{A}}\,A^*B,\qquad\mbox{where}\quad
A\in\End\cotanspace,\: B\in\End\tanspace.
\]

Further, the cases of algebras $\mathcal{A}$ of exterior ($\star=\wedge$)
and symmetric ($\star=\cdot$) forms will be used.

The phase space $\Phi$ is a contact manifold equipped with the
structure form \cite{LM}
\[
\contactForm=d\entr-\temp^{-1}d\intenergy+
\temp^{-1}\stress\cdot d\dfrm.
\]
The first law of thermodynamics means that
the \textit{thermodynamic state}
is a maximal integral manifold of this
differential form, i.~e. a Legendrian
manifold $L\subset (\Phi,\contactForm)$
of dimension $\dim (\End\tanspace)+1$.

Because the entropy is not involved
in the conservation laws
that govern medium motion, we eliminate it
from our geometrical description of the
thermodynamics.
To this end, we consider the projection $\phi$ of the
contact manifold $\contactSpace$ on the symplectic manifold
$(\symplecticSpace, d\contactForm)$,
where
$\symplecticSpace=\reals^{2}\times
\End\cotanspace\times\End\tanspace$
and
$\phi(\entr,\temp,\intenergy,\stress,\dfrm)=
(\temp,\intenergy,\stress,\dfrm)$.

Then the restriction of the mapping $\phi$ on the Legendrian
manifold $L$ is a local diffeomorphism on
the image $\widetilde{L}=\phi(L)$, and,
therefore, $\widetilde{L}\subset\widetilde{\Phi}$ is an immersed
Lagrangian manifold in a $2(\dim(\End T)+1)$-dimensional
symplectic manifold equipped with the structure form 
\[
d\contactForm =\temp^{-2}\left(d\temp\wedge d\intenergy+
\temp d\stress\wedge d\dfrm +\stress d\dfrm\wedge d\temp\right).
\]

Thus, the first law of thermodynamics can be reformulated
by saying that the thermodynamic state is
a Lagrangian submanifold of the symplectic manifold
$(\widetilde{\Phi},d\contactForm)$.

Given a Lagrangian manifold $\widetilde{L}$
the corresponding Legendrian manifold $L$
can be reconstructed up to the translation
along $\entr$-axis. It fits to our understanding
of entropy as a function defined up to a constant.

In addition, we also require (see \cite{LM} for more details)
that the quadratic differential form 
\[
\kappa =\temp^{-2}\left(d\temp\cdot d\intenergy+\temp
d\stress\cdot d\dfrm+\stress\cdot d\dfrm\cdot d\temp\right)
\]
be negative-definite on the Lagrangian submanifold $\widetilde{L}$.
The submanifold $\Sigma\subset\widetilde{L}$,
where quadratic form $\kappa$
is degenerated, divides $\widetilde{L}$ into domains,
where $\kappa$ is negative definite, and the rest. The domains,
where $\kappa$ is negative definite,
correspond to different phases of the
medium, all other domains are domains, where
the thermodynamic state is unstable.

The symplectic structure defines the Poisson
bracket on functions on $\symplecticSpace$ of the form
\[
\lbrack F,G\rbrack=
\frac{\temp}{2}
\left(
\frac{\partial G}{\partial\dfrm}\cdot
\frac{\partial G}{\partial\stress}
-\frac{\partial F}{\partial\dfrm}\cdot
\frac{\partial G}{\partial\stress}+
\temp\left(
\frac{\partial G}{\partial\intenergy}
\frac{\partial F}{\partial\theta}-
\frac{\partial G}{\partial\intenergy}
\frac{\partial G}{\partial\temp}
\right)-\stress\cdot
\left( 
\frac{\partial F}{\partial\intenergy}
\frac{\partial G}{\partial\stress}-
\frac{\partial G}{\partial\intenergy}
\frac{\partial G}{\partial\stress}
\right)\right).
\]

Hence, the thermodynamic state of the medium can be also defined by
equations 
\begin{equation}\label{lagEq}
F_{k}(\temp,\intenergy,\stress,\dfrm)=0,\quad
k=1,\ldots,\dim\left(\End T\right)+1,
\end{equation}
where $\lbrack F_{k},F_{l}\rbrack=0$
due to the system \eqref{lagEq}.

For example, if we have coordinates $(\intenergy,\dfrm)$ on 
$\widetilde{L}$, then the Legendrian manifold $L$ may be written in the form
\begin{equation}\label{eq:q-form}
\entr=f\left(\intenergy,\dfrm\right),\quad
\temp^{-1}=f_{\intenergy},\quad
\sigma=\left(f_{\intenergy}\right)^{-1}f_{\dfrm}.  
\end{equation}
Another presentation that we'll use
in this paper is given by the coordinates
$(\temp,\dfrm)$ on $\widetilde{L}$. Then
\[
\entr=\entr\left(\temp,\dfrm\right),\quad
\sigma=\sigma\left(\temp,\dfrm\right),\quad
\intenergy=\intenergy\left(\temp,\dfrm\right)
\]
on $\widetilde{L}$ the condition gives us 
\[
\left.\contactForm\right\vert_{\widetilde{L}}=
\left(\entr_{\temp}-\temp^{-1}\intenergy_{\temp}\right)
d\temp+\left(\entr_{\dfrm}-
\temp^{-1}\intenergy_{\dfrm}+\temp^{-1}\stress\right)d\dfrm=0,
\]
or 
\[
\intenergy_{\temp}=\temp\entr_{\temp},\quad
\intenergy_{\dfrm}=\temp\entr_{\dfrm}+\stress.
\]
Let us introduce the density of Helmholtz free energy
$\helm=\helm\left(\temp,\dfrm\right)$
on $L$, $\helm=\intenergy-\temp\entr$,
then we get the following equations for
the Lagrangian manifold $\widetilde{L}$ 
\begin{equation}\label{eq:qT-form}
\stress=\helm_{\dfrm},\quad
\intenergy=\left(\temp\helm\right)_{\temp}.
\end{equation}
The quadratic differential form $\kappa$ is of the form
\[
\kappa =-\sum_{i,j,k,l}f_{\dfrm_{ij}\dfrm_{kl}}
d\dfrm_{ij}\otimes d\dfrm_{kl}
+2f_{\intenergy\dfrm}\cdot d\intenergy\cdot d\dfrm
-f_{\intenergy\intenergy}d\intenergy^{2}
\]
in the first case, and 
\[
\kappa=\temp^{-2}\left(
(2\helm_{\temp}+\temp\helm_{\temp\temp})d\temp^2+
2(\helm_{\dfrm}+\temp\helm_{\dfrm\temp})d\temp d\dfrm+
\temp\sum_{i,j,k,l}h_{\dfrm_{ij}\dfrm_{kl}}
d\dfrm_{ij}\otimes d\dfrm_{kl}
\right)
\]
in the second one.

\section{Thermodynamic invariants of Newtonian media}

Let us consider a medium that
possesses a symmetry given by
an algebraic group $G\subset GL(T)$. Then
this $G$-action on the tangent
space $T$ can be prolonged to
the contact $G$-action on thermodynamic
phase space $\contactSpace$, if we
assume that this action
is trivial on
$\reals^3=(\entr,\temp,\intenergy)$
and natural on $\End\cotanspace\times\End\tanspace$.

Let $J_{1},\ldots,J_{N}$ be a set
of algebraically independent
rational $G$-invariants on $\Phi$, which
generate the field of rational $G$-invariants
and, therefore, separate regular $G$-orbits
(Rosenlicht theorem \cite{Ros}).

Then a
regular $G$-invariant thermodynamic
state, i.~e. a $G$-invariant algebraic
Legendrian
manifold $L\subset\Phi$ such that
almost all $G$-orbits in $L$ are regular,
can be written in the form of
$\entr=f(J_{1},\ldots,J_{N})$
in the case \eqref{eq:q-form}, or
$\helm=\helm(J_{1},\ldots,J_N)$
in the case \eqref{eq:qT-form},
where $f$ and $h$ are rational functions.

In this section, we consider Newtonian media,
i.~e. media with
a symmetry group
$G=O(\metric)\subset GL\left( T\right)$,
where $T$ is a Euclidean
vector space with a metric $\metric$,
$(x,y)\overset{\text{def}}{=}\metric(x,y)$.

To this end we study,
$O(\metric)$-orbits and $O(\metric)$-invariants
of the natural $O(\metric)$-action on $\End T$ (see \cite{Proc}).

Let $A\in\End\tanspace$ be a linear operator
and  $\cnj{A}\in\End\tanspace$
its adjoint operator with respect to
the metric $\metric$.

\begin{theorem}[Procesi \cite{Proc}]
	Algebra of polynomial $O(\metric)$-invariants on
	$A\in\End\tanspace$
	are generated by the Artin-Procesi invariants
	\[
		\mathcal{P}_{\alpha,\beta}(A)=
		\Tr\,(
		A^{\alpha_1}
		A^{\prime\beta_1}\cdots
		A^{\alpha_m}A^{\prime\beta_m}), \quad
		\sum_i(\alpha_i+\beta_i)\leq 2^n-1,
	\]
	where $\alpha=(\alpha_1,\ldots,\alpha_m)$,
	$\beta=(\beta_1,\ldots,\beta_m)$ are multi-indices.
\end{theorem}

The next theorem follows from the Procesi theorem,
the Rosenlicht theorem \cite{Ros} and the observation
that codimension of regular orbits equals
\[
\nu = n^2-\frac{n(n-1)}{2}=\frac{n(n+1)}{2}.
\]

\begin{theorem}
Field of rational invariant of the $O(\metric)$-action
on $\End\tanspace$ is generated by
any $\nu$ algebraically independent
Artin-Procesi invariants.
This field separates regular orbits.
\end{theorem}

Note that the Helmholtz free energy for Newtonian media
is $O(\metric)$-invariant because
of the trivial $O(\metric)$-action on
$(\entr,\intenergy,\temp)$, and therefore,
due to the Rosenlicht theorem,
$\helm=\helm(\temp,\mathcal{P}_{\alpha,\beta}(\dfrm))$
if $\helm$ is rational.

In this case, due to (\ref{eq:qT-form}),
we have the following state equation:
\[
\stress=
\parder{\helm}{\dfrm}=
\sum_{\alpha,\beta}\parder{\helm}{\mathcal{P}_{\alpha,\beta}}
\parder{\mathcal{P}_{\alpha,\beta}}{\dfrm}.
\]

If we consider media, which
satisfy `Hooke's law',
the Helmholtz free energy is
a quadratic function of $\dfrm$
and, therefore, has the form:
\[
\helm=
\frac{1}{2}\left(
a(\theta)\mathcal{P}_{2}(\dfrm)+
b(\theta)\mathcal{P}_{11}(\dfrm)+
c(\theta)\mathcal{P}_{1}^2(\dfrm)
\right)+
d(\theta)\mathcal{P}_{1}(\dfrm),
\]
where $a,b,c,d$ are some functions.

In this case the state equations take the form:
\[
\stress =
a(\temp)\cnj{\dfrm}+
b(\temp)\dfrm+
(c(\temp) \Tr\dfrm +d(\temp))\idop.
\]

We call the functions $a,b,c$
viscosities, and $-d$ pressure,
though they are not completely adequate to
their names.

Usually, the cases when the operator
$\dfrm$ is self-adjoint are considered, in such cases
there are only two viscosities.

\section{Thermodynamic invariants
of Newtonian media with inner structure}

Let a Euclidean vector space $(T,g)$ be the orthogonal
direct sum of a vertical $(V,g_F)$
and a horizontal $(H,g_{B})$ Euclidean spaces, that is
\begin{equation}\label{eq:split}
(T,g)=\left( V,g_{F}\right) \oplus (H,g_{B}),  
\end{equation}
where $\dim V=m$, $\dim H=n$.

In this section we study invariants
of linear operators $\End T$ equipped with
the natural $O(g_{F})\times O(g_{B})$-action.

Let $\Pi_V$ be orthogonal projector on $V$.

\begin{theorem}
	Algebra of polynomial $O(g_{F})\times O(g_{B})$-invariants
	on
	$A\in\End\tanspace$
	is generated by Artin-Procesi invariants
	\[
		\mathcal{P}_{\alpha,\epsilon,\beta}(A)=
		\Tr\,(
		A^{\alpha_1}\Pi_V^{\epsilon_1}A^{\prime\beta_1}\cdots
		A^{\alpha_k}\Pi_V^{\epsilon_k}A^{\prime\beta_k}),
		\quad
		\sum_i(\alpha_i+\epsilon_i+\beta_i)\leq 2^{n+m}-1,
	\]
	where $\alpha=(\alpha_1,\ldots,\alpha_m)$,
	$\epsilon=(\epsilon_1,\ldots,\epsilon_m)$,
	$\beta=(\beta_1,\ldots,\beta_m)$ are multi-indices.
\end{theorem}

It follows from the Rosenlicht theorem that
we should expect
the same number of generators in
the field of rational $O(g_{F})\times O(g_{B})$-invariants,
then dimension
of a general orbit equals
\[
(n+m)^2-\frac{n(n-1)}{2}-\frac{m(m-1)}{2}=
\frac{n(n+1)}{2}+\frac{m(m+1)}{2}+2mn.
\]

Similar to the ordinary Newtonian media, for
the Newtonian media with inner structure,
we have the following state equations:
\[
	\parder{\helm}{\dfrm}=
	\sum_{\alpha,\epsilon,\beta}
	\parder{\helm}{\mathcal{P}_{\alpha,\epsilon,\beta}}
	\parder{\mathcal{P}_{\alpha,\epsilon,\beta}}{\dfrm}.
\]

In the case when the media satisfy `Hooke's law',
the Helmholtz free energy is
a quadratic function of $\dfrm$ and therefore has the form:
\begin{align*}
\helm=
\frac{1}{2}\left(
a_1(\theta)\Tr\,\dfrm^2+
a_2(\theta)\Tr\,(\dfrm\cnj{\dfrm})+
a_3(\theta)\Tr^2\,\dfrm+a_4(\theta)\Tr^2\,(\dfrm\Pi_V)+\right.\\
\left.a_5(\theta)\Tr\,(\cnj{\dfrm}\dfrm\Pi_V)+
a_6(\theta)\Tr\,(\dfrm\cnj{\dfrm}\Pi_V)
\right)+
b_1(\theta)\Tr\,\dfrm+
b_2(\theta)\Tr\,\dfrm\Pi_V,
\end{align*}
where $a_1,\ldots,a_6,b_1,b_2$ are some functions.

In this case the state equations take the form:
\begin{align*}
\stress =
&\,a_1(\temp)\cnj{\dfrm}+
a_2(\temp)\dfrm+
(a_3(\temp)(\Tr\,\dfrm)+b_1(\temp))\idop+\\
&(a_4(\temp)\Tr\,(\dfrm\Pi_V)+b_2(\temp))\Pi_V+
a_5(\temp)\dfrm\Pi_V+
a_6(\temp)\Pi_V\dfrm.
\end{align*}

\section{Conservation laws and motion of media}

\subsection{Preliminaries}

The general laws of conservation of mass,
momentum and energy are used to
describe the motion of media.
This can be done on an arbitrary Riemannian
manifold $(\mnfld,\metric)$, the only difference
between media with or without inner structure
is in the description of a flow velocity field 
$\flowvel$ and the thermodynamics of media. Thus,
for media without inner structure
we have no extra conditions for vector fields
and thermodynamics but in
presence of inner structures we have to specify them.

Thus, in this section we consider a Riemannian manifold
$(\mnfld,\metric)$
and write down the conservation laws
for an arbitrary vector field $\flowvel$ on $\mnfld$.
We'll assume that $\mnfld$ is an oriented manifold and
$\Omega=\Omega_{\metric}$ is
the volume form associated with the metric $\metric$.
We denote by
$\nabla$ and $\covdif$ the Levi-Civita
connection and the covariant differential also
associated with metric $\metric$.

By 
\[\matD
=\frac{\partial}{\partial t}+\nabla_{X},
\]
we denote the material derivative.

\subsection{Divergence operator}

The divergence of a vector field, $\diver X$, is defined by
the standard way through Lie derivative:
\[
\LieDerivative{\flowvel}{\Omega}=(\diver\flowvel)\,\Omega.
\]
On the other hand, the covariant differential
$\covdif\flowvel\in T^{*}\otimes T$ is a field
of linear operators acting in the tangent spaces
and we get
\begin{equation}\label{eq:divX}
\diver\flowvel =\Tr(\covdif\flowvel).
\end{equation}
To see that we get an equivalent construction,
let us write down the latter in
local coordinates $x_1,\ldots,x_n$.

We get 
\[
d_{\nabla}X=
\sum_{i,j}\left( \parder{X_i}{x_j}+
\sum_k \Gamma_{kj}^{i}X_k\right)
\frac{\partial }{\partial x_i}\otimes dx_j,
\]
where $X=\sum X_i\frac{\partial }{\partial x_i}$,
and $\Gamma_{kj}^{i}$ are the Christoffel symbols.

Taking now $(x_1,\ldots,x_n)$
to be the normal coordinates we
get equality \eqref{eq:divX}.

This observation allows us to extend
the divergence operator on other
tensors. The more important case for us
is the case of linear
operators $A\in\End T=T\otimes T^*$.

In this case, $\covdif A\in T^{*}\otimes \End T
=T^*\otimes T\otimes T^*$ and,
by taking $(1,2)$-contraction $c_{1,2}$, we get
a differential $1$-form that we call divergence
of operator $A$:
\[
\diver A=c_{1,2}\left( d_{\nabla }A\right) \in T^*.
\]
In local coordinates we have 
\begin{align*}
&A=\sum a_i^k\frac{\partial }{\partial x_i}\otimes dx_k,\\
&\covdif\left( \frac{\partial }{\partial x_i}\right) =
\sum \Gamma_{ij}^k
\frac{\partial}{\partial x_k}\otimes dx_j,\\
&\covdif(dx_i) =-\sum\Gamma_{jk}^{i}dx_k\otimes dx_j
\end{align*}
and, therefore,
\[
\diver A=\sum_{i,k}\left(\parder{a_{i}^{k}}{x_{i}}
+\sum_{j}(a_{j}^{k}\Gamma_{ij}^{i}-a_{i}^{j}\Gamma _{ik}^{j})
\right) dx_{k}.
\]

Let $A=X\otimes \omega $, where $X$ is
a vector field and $\omega $ is a
differential $1$-form. Then it is easy to check that 
\begin{equation}\label{usForm}
\diver\left( X\otimes \omega \right) =(\diver X)\,\omega
+\nabla_{X}\omega.
\end{equation}

We consider the stress tensor $\stress\in\End T$ as
the surface force
$\widehat{\stress}=g(\stress( \nu)
,\cdot) $ $ \in \cotanspace$ applied to an imaginary surface
orthogonal to a normal vector $\nu$.
In our case we cannot directly find the `integral
sum' of all forces applied to a volume,
since each of the `applied forces'
belongs to a different space.

Take a small volume $\Delta V$ with a border $S$
around a center point $a\in M$ and consider
parallel transports $\gamma_{a,x}:T_{a}\rightarrow T_{x}$
along the shortest geodesic line connected points $x$ and $a$.

Let $f(x)=\gamma_{a,x}^{*}\left(\widehat{\stress}\right)
\in T_{a}^{*}$ be the images of the surface forces.
Then the force applied to the volume we'll be understand
the integral $\int_{S}f\,ds$,
where $f$ considered as a vector
valued function $f:\Delta V\rightarrow T_{a}^{*}$

To see that the density of this
force equals $\diver\stress$ let take
normal coordinates $(x_1,\ldots,x_n)$
centered at the point $a$.
Then $g_{ij}=\delta_{ij}+o_{2}(x)$,
$\Gamma_{ij}^{k}\in o_{1}(x)$, where
we denoted by $o_{l}(x)$
functions having zero of order $l$
at the point $a$, and therefore, due to
the above formula,
$\diver\stress=\sum\sigma_{i,i}^{k}(a)dx_k$
at the point $a$.

On the other hand,
$f(x)=\sum\left(\sigma_{i}^{k}(x)+o_{1}(x) \right) \nu_{i}$
and, therefore,
\[
\int_{S}f\,ds=\int_{S}\sum \stress_{i}^{k}(x)
\nu_{i}ds+o_{1}\left( \Delta V\right)
=\left( \diver_{a}\sigma \right) 
\Delta V+o_{1}\left( \Delta V\right),
\]
i.~e. the density of internal force is $-\diver\stress$.

\subsubsection{Conservation of mass}

The idea that the amount of a medium
contained in an elementary volume is
conserved along the flow of the medium
can be expressed with the equation
\[
\left( \frac{\partial }{\partial t}+
\LieDerivative{\flowvel}{\!}\right) \left(\dens\,\Omega\right) =0,
\]
or equally
\begin{equation}\label{eq:mass}
\frac{d\dens}{dt}+\dens\,\diver\flowvel = 0.
\end{equation}

\subsubsection{Conservation of momentum,
or the Navier--Stokes equation of motion}

Considering the metric $g$ as the isomorphism
$g:T\rightarrow T^*$,
denote by $X^{\flat }=g(X)\in T^*$
the differential $1$-form dual to
a vector field $X$, and by $\alpha^{\flat}$
the vector field dual to
a differential $1$-form $\alpha$.

Then a differential form $\omega =\rho X^{\flat}$
may be considered as the
momentum density of the media.

Newton's second law states that
the force equals to the mass multiplied
by the acceleration. In our case,
$\dfrac{dX}{dt}$ is the acceleration and,
therefore, the law states that
\begin{equation}\label{eq:2ndNewton}
\dens \frac{d\flowvel}{dt}=-\diver^{\flat}\stress.
\end{equation}

The Levi-Civita connection preserves the metric $\metric$.
Therefore, due to the
above formula, we have 
\[
\rho \frac{dX^{\flat }}{dt}=-\diver\stress,
\]
and due to (\ref{usForm}) we get 
\[
\parder{\omega}{t}+\diver( \flowvel\otimes \omega)
=-\diver\stress.
\]
The last relation in the form 
\begin{equation}\label{MomConserv}
\parder{\omega}{t}+
\diver( \flowvel\otimes \omega +\stress)=0,
\end{equation}
is called the equation of the momentum conservation.

\subsubsection{Conservation of energy}

The law of energy conservation is written as follows
\begin{equation}\label{eq:energy1}
\parder{\totenergy}{t}=-\diver \energyFlux,
\end{equation}
where $\energyFlux$ is the total energy flux vector.

This vector is the sum of
a convective term $\totenergy\flowvel$,
a mechanical energy flux $\stress(\flowvel)$
and a heat flow $\mathcal{J}_q$ (for details, see \cite{deG}):
\[
\mathcal{J}_e=\totenergy\flowvel + \stress(\flowvel)
\heatFlow
\]

Substitution the velocity field $\flowvel$
into \eqref{eq:2ndNewton}
leads to the kinetic energy balance equation
\begin{equation}\label{eq:kinetic}
\dens
\matD{\frac{\inner{\flowvel}{\flowvel}}{2}}
= -(\diver\stress)(\flowvel)+
\inner{{\stress}}{\dfrm}
\end{equation}

Recall that the total energy enclosed in
an elementary volume is the sum
of kinetic, potential and internal energy
\[
\totenergy = \frac{\dens\langle\flowvel,\flowvel\rangle}{2}
+\intenergy.
\]

Using this identity and the equations \eqref{eq:mass},
\eqref{eq:kinetic} we get
\[
\matD{\intenergy}+\intenergy\diver\flowvel+\diver(\heatFlow)+
\langle\stress,\dfrm\rangle=0.
\]

Commonly, the heat flow is given by Fourier's law 
\[
\heatFlow=-\thermcond(\grad T),
\]
where $\thermcond\in\End\tanspace$ is the thermal conductivity of the medium.

Summarizing we have the following system of differential equations
describing media with inner structures:
\[
\left\{
\begin{aligned}
&\frac{d\dens}{dt}+\dens\diver\flowvel=0,\\
&\dens \frac{d\flowvel}{dt}=-\diver^{\flat}\stress,\\
&\frac{d\intenergy}{dt}+\intenergy\diver\flowvel+\diver(\heatFlow)+
\langle\stress,\dfrm\rangle=0,
\end{aligned}
\right.
\]
where 
\begin{align*}
&\sigma =\frac{\partial h}{\partial \Delta },\ \ \varepsilon =h+\theta 
\frac{\partial h}{\partial \theta }, \\
&X\ \text{is a }\pi \text{-projectable vector field.}
\end{align*}

\section*{Acknowledgements}

All three authors are partially supported by the Russian Foundation
for Basic Research, Grant 18-29-10013.

Valentin Lychagin and Sergey Tychkov are partially supported by the
Foundation for the Advancement of
Theoretical Physics and Mathematics ``BASIS'', Grant 19-7-1-13-2.


\begin{thebibliography}{9}
\bibitem{B} Batchelor, G. K. An introduction to fluid dynamics.  Cambridge
Mathematical Library. Cambridge University Press, Cambridge, 1999. xviii+615
pp.

\bibitem{GLTR} Gorinov, Anton A.; Lychagin, Valentin V.; Roop, Mikhail D.;
Tychkov, Sergey N. Gas flow with phase transitions: thermodynamics and the
Navier-Stokes equations. Nonlinear PDEs, their geometry, and applications,
209--222, Tutor. Sch. Workshops Math. Sci., Springer, 2019

\bibitem{DLT1} Duyunova, A. A.; Lychagin, V. V.; Tychkov, S. N.
Classification of the thermodynamic state equations of a viscous fluid.
(Russian) Dokl. Akad. Nauk 473 (2017), no. 6, 635--639; translation in Dokl.
Math. 95 (2017), no. 2, 172--175

\bibitem{DLT2}  Duyunova, Anna; Lychagin, Valentin; Tychkov, Sergey
Differential invariants for spherical layer flows of viscid fluids. J. Geom.
Phys. 130 (2018), 288--292.

\bibitem{Kr} Kruglikov Boris, Topological classification of Leggett systems
in an integrable case for 3He--A, Russ. Math. Surv., 46 (4), pp.
179--181,1991.

\bibitem{KrL} Kruglikov Boris, Lychagin Valentin, Global Lie-Tresse theorem,
Selecta Math. (NS) 22 (3) (2016) 1357--1411.

\bibitem{LM} Lychagin, Valentin, Contact geometry, measurement, and
thermodynamics. Nonlinear PDEs, their geometry, and applications, 3--52,
Tutor. Sch. Workshops Math. Sci., Springer, 2019.

\bibitem{LR} Lychagin, Valentin; Roop, Mikhail; Real gas flows issued from a
source. Anal. Math. Phys. 10 (2020), no. 1.

\bibitem{Ros} Rosenlicht M., A remark on quotient spaces,
An. Acad. Brasil.
Cienc. 35 (1963) 487--489.

\bibitem{Proc} Procesi C., Lie groups: an approach through
invariants and representations, Springer, (2005).

\bibitem{Vard} Vardoulakis, Ioannis.
Cosserat Continuum Mechanics:
With Applications to Granular Media. Vol. 87. Springer, 2018.

\bibitem{Cos} Cosserat, Eugene, and Fran\c cois Cosserat.
Th\`eorie des corps d\'eformables. (1909).

\bibitem{Sp1} Spencer A. J. M.; Rivlin R. S.
Further results in the theory of matrix polynomials.
Arch. Rational Mech. Anal. 4 (1960), 214--230.

\bibitem{Sp2} Spencer, A. J. M. Theory Invariants,
in Continuum Mechanics. vol.1, Longman, (1980).


\bibitem{deG}De Groot, Sybren Ruurds,
and Peter Mazur. ``Non-equilibrium thermodynamics.'' (2013).

\end{thebibliography}
\end{document}